\documentclass[aps,prd,groupedaddress,nofootinbib,superscriptaddress,tighten,preprint]{revtex4}
\usepackage{mathbbold}
\usepackage{amsfonts}
\usepackage{amssymb}
\usepackage{amsmath,color}
\usepackage{epsfig}
\usepackage{graphicx,epsf,epsfig,subfigure}
\usepackage{bbm}
\def\slc#1{\setbox0=\hbox{$#1$}           
    \dimen0=\wd0                                 
    \setbox1=\hbox{/} \dimen1=\wd1               
    \ifdim\dimen0>\dimen1                        
       \rlap{\hbox to \dimen0{\hfil/\hfil}}      
       #1                                        
    \else                                        
       \rlap{\hbox to \dimen1{\hfil$#1$\hfil}}   
       /                                         
    \fi}

\begin{document}

\title{Nonstandard interaction effects on neutrino parameters at medium-baseline reactor antineutrino experiments}

\author{Tommy Ohlsson}

\email{tohlsson@kth.se}

\affiliation{Department of Theoretical Physics, School of
Engineering Sciences, KTH Royal Institute of Technology, AlbaNova
University Center, 106 91 Stockholm, Sweden}

\author{He Zhang}

\email{he.zhang@mpi-hd.mpg.de}

\affiliation{Max-Planck-Institut f\"{u}r Kernphysik, Saupfercheckweg
1, 69117 Heidelberg, Germany}

\author{Shun Zhou}
\email{shunzhou@kth.se}

\affiliation{Department of Theoretical Physics, School of
Engineering Sciences, KTH Royal Institute of Technology, AlbaNova
University Center, 106 91 Stockholm, Sweden}

\begin{abstract}
Precision measurements of leptonic mixing parameters and the determination of the neutrino mass hierarchy are the primary goals of the forthcoming medium-baseline reactor antineutrino experiments, such as JUNO and RENO-50. In this work, we investigate the impact of nonstandard neutrino interactions (NSIs) on the measurements of $\{\sin^2 \theta^{}_{12}, \Delta m^2_{21}\}$ and $\{\sin^2 \theta^{}_{13}, \Delta m^2_{31}\}$, and on the sensitivity to the neutrino mass hierarchy, at the medium-baseline reactor experiments by assuming a typical experimental setup. It turns out that the true mixing parameter $\sin^2 \theta^{}_{12}$ can be excluded at a more than $3\sigma$ level if the NSI parameter $\varepsilon^{}_{e\mu}$ or $\varepsilon^{}_{e\tau}$ is as large as $2\%$ in the most optimistic case. However, the discovery reach of NSI effects has been found to be small, and depends crucially on the CP-violating phases. Finally, we show that NSI effects could enhance or reduce the discrimination power of the JUNO and RENO-50 experiments between the normal and inverted neutrino mass hierarchies.
\end{abstract}


\maketitle


\section{introduction}
\label{sec:intro}

The experimental endeavor in the past decades has established that the phenomenon of neutrino flavor transitions is described by neutrino oscillations at leading order. Now that the smallest leptonic mixing angle $\theta_{13}$ has been measured very accurately at reactor~\cite{An:2012eh,Ahn:2012nd,Abe:2012tg,An:2012bu} and accelerator~\cite{Abe:2013xua} neutrino experiments, one of the next major and open problems is the discrimination between normal and inverted neutrino mass hierarchies. Towards this challenge, different types of experiments have been suggested using long-baseline accelerator-based neutrinos, atmospheric neutrinos, supernova neutrinos, or reactor neutrinos. Recent studies indicate that a dedicated medium-baseline reactor antineutrino experiment with sufficient statistics and unprecedented detector performance provides an opportunity to determine the neutrino mass hierarchy and probe with high precision the other neutrino parameters (see, e.g., Ref.~\cite{Kettell:2013eos} and references therein). In the near future, potential projects along this direction include the JUNO~\cite{Li:2013zyd} and RENO-50~\cite{RENO50} experiments.

Beyond the standard oscillation picture, new physics may appear in future neutrino experiments in the form of unknown couplings involving neutrinos, which are usually referred to as nonstandard neutrino interactions~\cite{Wolfenstein:1977ue,Valle:1987gv,Guzzo:1991hi,Roulet:1991sm,Grossman:1995wx} (NSIs). In fact, NSIs are predicted as dimension-six and higher-order operators in many interesting extensions of the Standard Model, e.g., $R$-parity violating supersymmetric theories, left-right symmetric models, grand unification theories, extra dimensions, and various seesaw models (see, e.g., the recent review on NSIs~\cite{Ohlsson:2012kf} and references therein). Basically, all modern extensions could give rise to NSIs, and the investigation of NSIs could be very helpful in revealing additional new physics behind neutrino flavor transitions. In addition, it plays an important complementary role to direct searches of new physics at colliders. NSI effects at reactor antineutrino experiments have been studied with emphasis on mimicking effects~\cite{Ohlsson:2008gx}, the discrepancy between source and detector NSIs~\cite{Leitner:2011aa}, comparisons between reactor and accelerator experiments~\cite{Kopp:2007ne,Adhikari:2012vc}, and modifications of event rates and the impact on the measurements of standard oscillation parameters~\cite{Khan:2013hva}. On the other hand, the model-independent bounds on production and detection NSIs have been derived in Ref.~\cite{Biggio:2009nt}, and all bounds are basically of order $10^{-2}$.

In this work, we investigate NSI effects at a typical medium-baseline reactor antineutrino experiment, in particular, the mimicking effects in the precision measurements of neutrino parameters, the discovery reach of NSI effects, and the distortion of the neutrino mass hierarchy determination. Specifically, we concentrate on an experimental setup similar to the JUNO experiment and show that, for five years of running, a clear hint for nonstandard neutrino physics can be provided. The remainder of this Letter is organized as follows. In Section~\ref{sec:formalism}, we present the general formalism and derive relevant antineutrino survival probability formulas used in the subsequent analysis. Illustrations of NSI effects on the corresponding oscillation probabilities and energy distributions of neutrino events are also given. In Section~\ref{sec:numerics}, a detailed numerical analysis of the observability of the NSI effects at JUNO is performed. Finally, in Section~\ref{sec:summary}, we summarize and state our conclusions.

\section{Basic Formalism}
\label{sec:formalism}

In this section, we present the general formulas for reactor antineutrino oscillations with NSIs. For a realistic experiment, NSIs may appear both in the production and detection processes, and the neutrino states produced in the source and observed at the detector can be treated as
superpositions of pure orthonormal flavor states:
\begin{eqnarray}\label{eq:normalization}
|\hat \nu^{\rm s}_\alpha \rangle & = & \frac{1}{N^{\rm s}_\alpha} \left( |\nu_\alpha \rangle + \sum_{\beta=e,\mu,\tau} \varepsilon^{\rm s}_{\alpha\beta} |\nu_\beta\rangle  \right) \ , \\
\langle \hat  \nu^{\rm d}_\beta| & = & \frac{1}{N^{\rm d}_\beta} \left( \langle \nu_\beta | + \sum_{\alpha=e,\mu,\tau}
\varepsilon^{\rm d}_{\alpha \beta} \langle  \nu_\alpha  | \right) \ ,
\end{eqnarray}
where the superscripts `s' and `d' denote the source and the detector, respectively, with the normalization factors\footnote{In the calculation of  the number of events in detectors, the normalization factors are canceled with the NSI factors in charged-current cross sections. However, for reactor antineutrino experiments, the neutrino fluxes are extracted from the measurement instead of a Monte Carlo simulation. Hence, the normalization factors should be taken into account. See also Ref.~\cite{Antusch:2006vwa} for a detailed discussion.}
\begin{eqnarray}\label{eq:factor}
N^{\rm s}_\alpha & = & \sqrt{\left[\left(\mathbbold{1} + \varepsilon^{\rm s}
\right)\left(\mathbbold{1} + \varepsilon^{{\rm s} \dagger} \right)
\right]_{\alpha\alpha}} \;~~ , \\
N^{\rm d}_\beta & = & \sqrt{\left[\left(\mathbbold{1} + \varepsilon^{{\rm d}
\dagger} \right)\left(\mathbbold{1} + {\varepsilon^{\rm d}} \right)
\right]_{\beta \beta}}   \;~~ .
\end{eqnarray}

In general, the NSI parameter matrices $\varepsilon^{\rm s}$ and $\varepsilon^{\rm d}$ are arbitrary and nonunitary, indicating that neither $|\hat  \nu^{\rm s}_\alpha \rangle $ nor $\langle \hat \nu^{\rm d}_\beta|$ are orthonormal states. As first observed in Ref.~\cite{Schechter:1980gr}, heavy neutrino states responsible for neutrino mass generation decouple from oscillation processes, so that the unitarity of the leptonic mixing matrix is slightly violated. The nonunitary effects, which could be significant in low-scale seesaw models~\cite{Mohapatra:1986bd,Pilaftsis:1991ug}, can be regarded as one type of NSIs with the requirement $\varepsilon^{\rm s} = \varepsilon^{{\rm d} \dagger}$ \cite{Meloni:2009cg}. For reactor antineutrinos, the leading-order NSIs are of the $V\pm A$ type as long as CPT is conserved, and it is very common to assume $\varepsilon^{\rm s}_{e\alpha} = \varepsilon^{{\rm d} *}_{\alpha e}$~\cite{Kopp:2007ne}. Thus, we will take $\varepsilon^{\rm s}_{e\alpha} = \varepsilon^{{\rm d}*}_{\alpha e} = \varepsilon_{e\alpha}{\rm e}^{{\rm i}\phi_{e\alpha}}$ (with $\varepsilon_{e\alpha}$ being the modulus of $\varepsilon^{\rm s}_{e\alpha}$) in the current consideration and neglect the superscripts `s' and `d' throughout the following parts of this work. Furthermore, the typical energy of antineutrinos produced in nuclear reactors is around a few MeV, which indicates that Earth matter effects are very small and can safely be neglected in most situations. Hence, the propagation of neutrino flavor states is governed by the vacuum Hamiltonian
\begin{eqnarray}\label{eq:Hamiltonian}
{\cal H} = \frac{1}{2E} U \cdot {\rm diag} (m^2_1,m^2_2,m^2_3) \cdot U^\dagger \ ,
\end{eqnarray}
where the leptonic mixing matrix $U$ is usually parametrized in the standard form by using three mixing angles and one CP-violating phase
\begin{eqnarray}\label{eq:parametrization}
U = \left(
\begin{matrix}c_{12} c_{13} & s_{12} c_{13} & s_{13}
{\rm e}^{-{\rm i}\delta} \cr -s_{12} c_{23}-c_{12} s_{23}
s_{13}
{\rm e}^{{\rm i} \delta} & c_{12} c_{23}-s_{12} s_{23} s_{13}
{\rm e}^{{\rm i} \delta} & s_{23} c_{13} \cr
 s_{12} s_{23}-c_{12} c_{23} s_{13}
{\rm e}^{{\rm i} \delta} & -c_{12} s_{23}-s_{12} c_{23} s_{13}
{\rm e}^{{\rm i} \delta} & c_{23} c_{13}\end{matrix}
\right) \ ,
\end{eqnarray}
with $c_{ij} \equiv \cos \theta_{ij}$ and $s_{ij} \equiv \sin \theta_{ij}$ (for $ij=12$, $13$ and $23$).

Including the NSI effects, we arrive at the amplitude for the $\hat  \nu_e \rightarrow \hat  \nu_e$ oscillation channel
\begin{eqnarray}\label{eq:A1}
{\cal A}_{ee}(L) = \frac{1}{N^2_e}
\langle \hat  \nu_e |
\exp\left({-{\rm i} {\cal H} L}\right) | \hat  \nu_e \rangle = \frac{1}{N^2_e} \left( { A} + {\varepsilon} {
A} + { A} \varepsilon^\dagger + \varepsilon A \varepsilon^\dagger
\right)_{ee} \ ,
\end{eqnarray}
where $L$ is the propagation distance and the explicit form of $A$ is a coherent sum over the contributions of all mass eigenstates
$\nu_i$, i.e., ${A}_{\alpha\beta} = \sum_i U^*_{\alpha i} U_{\beta i} \exp\left(-{\rm i} m^2_i L/2E\right)$. Inserting ${A}_{\alpha\beta} $ into Eq.~\eqref{eq:A1}, one can rewrite the amplitude in a compact form
\begin{eqnarray}\label{eq:A2}
{\cal A}_{ee}(L) = \sum_i {\cal R}_i  \exp\left(-{\rm i}
\frac{m^2_i L}{2E}\right)
\end{eqnarray}
with
\begin{eqnarray}\label{eq:J}
{\cal R}_i & = &  \frac{|U_{ei}|^2 +  \displaystyle 2 \sum_\alpha {\rm Re} \left(U^*_{\alpha i} U_{e i} \varepsilon_{e\alpha} {\rm e}^{{\rm i} \phi_{e\alpha}}\right) + \displaystyle \sum_{\alpha, \beta}  {\rm Re} \left[ U^*_{\alpha i} U_{\beta i} \varepsilon_{e\alpha} \varepsilon_{e\beta} {\rm e}^{{\rm i} (\phi_{e\alpha} - \phi_{e\beta})}\right]}{1 + 2 \varepsilon_{ee} \cos \phi_{ee} + \varepsilon^2_{ee} + \varepsilon^2_{e\mu} + \varepsilon^2_{e\tau}} \nonumber \\
& = & |U_{ei}|^2 +  2 \sum_{\alpha \neq e} {\rm Re} \left(U^*_{\alpha i} U_{e i} \varepsilon_{e\alpha} {\rm e}^{{\rm i} \phi_{e\alpha}}\right) + {\cal O}(\varepsilon^2) \; .
\end{eqnarray}
It is a general feature from the first row of Eq.~(\ref{eq:J}) that ${\cal R}_i$ is real for the survival probabilities, which is actually a reflection of CPT conservation. Therefore, the amplitudes are identical for neutrino and antineutrino oscillations. However, this is no longer true when matter effects are non-negligible, since the ordinary Earth environment is CP asymmetric~\cite{Akhmedov:2001kd,Jacobson:2003wc}. Furthermore, in the limit $\varepsilon^{}_{e\alpha} \rightarrow 0$, we have ${\cal R}_i \to |U_{ei}|^2$, and hence, Eq.~\eqref{eq:A2} is reduced to the standard oscillation amplitude.

In practice, it is more useful to express the ${\cal R}_i$ parameters in terms of the leptonic mixing parameters, i.e.,
\begin{eqnarray}
{\cal R}_1 &=& c^2_{12} c^2_{13} - 2s_{12} c_{12} c_{13} \varepsilon_\phi - 2 c^2_{12} s_{13} c_{13} \varepsilon_\delta + {\cal O}(\varepsilon^2) \; , \\
{\cal R}_2 &=& s^2_{12} c^2_{13} + 2s_{12} c_{12} c_{13} \varepsilon_\phi - 2 s^2_{12} s_{13} c_{13} \varepsilon_\delta + {\cal O}(\varepsilon^2) \; , \\
{\cal R}_3 &=& s^2_{13} + 2 s_{13} c_{13} \varepsilon_\delta + {\cal O}(\varepsilon^2) \; ,
\end{eqnarray}
where the auxiliary parameters $\varepsilon_\phi$ and $\varepsilon_\delta$ are defined as
\begin{eqnarray}
\varepsilon_\phi &\equiv& c_{23} \varepsilon_{e\mu} \cos \phi_{e\mu} - s_{23} \varepsilon_{e\tau} \cos \phi_{e\tau} \; , \label{eq:epsphi} \\
\varepsilon_\delta &\equiv& s_{23} \varepsilon_{e\mu} \cos (\phi_{e\mu}-\delta) + c_{23} \varepsilon_{e\tau} \cos (\phi_{e\tau} - \delta) \; . \label{eq:epsdelta}
\end{eqnarray}
One observes that the $\varepsilon_{ee}$-dependent terms disappear from the survival probability, as a consequence of our chosen normalization. Moreover, if all CP-violating phases are vanishing, the auxiliary NSI parameters $\{\varepsilon_\phi, \varepsilon_\delta\}$ are related to the original ones $\{\varepsilon_{e\mu}, \varepsilon_{e\tau}\}$ by a rotation with the angle $\theta_{23}$.

Now, it is straightforward to derive the survival probability for electron antineutrinos
\begin{equation}\label{eq:P}
P(\bar{\nu}_e \to \bar{\nu}_e) = \sum_{i, j} {\cal R}_i {\cal R}_j - 4 \sum_{i > j} {\cal R}_i {\cal R}_j \sin^2 \frac{\Delta m^2_{ij} L}{4E}  \; ,
\end{equation}
where $\Delta m^2_{31} \equiv m^2_3 - m^2_1 > 0$ for normal neutrino mass hierarchy (NH) and $\Delta m^2_{31} < 0$ for inverted neutrino mass hierarchy (IH). Here and henceforth, the hat in the antineutrino state $| \bar{\hat \nu}_e \rangle$ will be omitted, hopefully without causing any confusion.

Then, considering a medium-baseline neutrino experiment, such as JUNO or RENO-50, we find that Eq.~\eqref{eq:P} approximates to
\begin{eqnarray}\label{eq:Papp}
P(\bar{\nu}_e \to \bar{\nu}_e) & \approx & 1 - 4s_{12} c_{12} c^3_{13}\left[s_{12} c_{12} c_{13} + 2 \left(\cos 2\theta_{12} \varepsilon_\phi - \sin 2\theta_{12} s_{13} \varepsilon_\delta\right)\right] \sin^2 \frac{\Delta m^2_{21} L}{4E} \nonumber \\
&&- 4 s_{13} c_{13} \left(s_{13} c_{13} c^2_{12} - \sin 2\theta_{12} s_{13} \varepsilon_\phi + 2 c^2_{12} \cos 2\theta_{13} \varepsilon_\delta \right) \sin^2 \frac{\Delta m^2_{31} L}{4E} \nonumber \\
&&- 4 s_{13} c_{13} \left(s_{13} c_{13} s^2_{12} + \sin 2\theta_{12} s_{13} \varepsilon_\phi + 2 s^2_{12} \cos 2\theta_{13} \varepsilon_\delta \right) \sin^2 \frac{\Delta m^2_{32} L}{4E}\; ,
\end{eqnarray}
where the higher-order terms ${\cal O}(\varepsilon^2)$ have been neglected. Note that Eq.~\eqref{eq:Papp} reproduces the standard survival probability in the limit $\varepsilon_\phi$, $\varepsilon_\delta \to 0$. It is worthwhile to note that $\varepsilon_\phi$ and $\varepsilon_\delta$ can be vanishingly small if, e.g., $\phi_{e\mu} = \phi_{e\tau} = \pm \pi/2$ and $\delta = 0$. In this case, only the higher-order terms ${\cal O}(\varepsilon^2)$ appear in the survival probability. In addition, the difference between the oscillation probability in the NH case and that in the IH case, i.e., $\Delta P \equiv P^{\rm NH}(\bar{\nu}_e \to \bar{\nu}_e) - P^{\rm IH}(\bar{\nu}_e \to \bar{\nu}_e)$, can be written as
\begin{equation}\label{eq:deltaP}
\Delta P = 4 s_{13} c_{13} \left(s_{13} c_{13} s^2_{12} + \sin 2\theta_{12} s_{13} \varepsilon_\phi + 2 s^2_{12} \cos 2\theta_{13} \varepsilon_\delta \right) \sin \frac{\Delta m^2_{21} L}{2E} \sin \frac{\Delta m^2_{31} L}{2E} \; ,
\end{equation}
indicating the impact of the NSI parameters on the discrimination between NH and IH.

\begin{figure}[t]
\begin{center}\vspace{0.cm}
\includegraphics[width=0.8\textwidth]{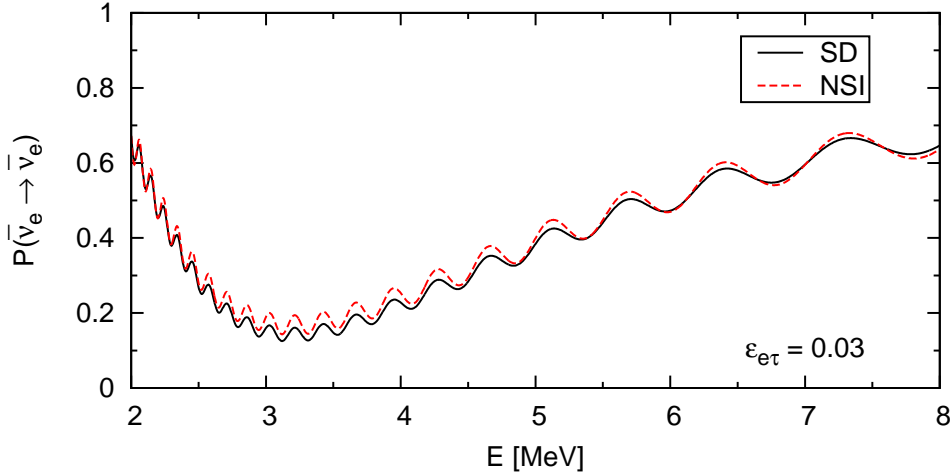}\vspace{-0.2cm}
\caption{\label{fig:Pee} The survival probability $P(\bar{\nu}_e \to \bar{\nu}_e)$ as a function of the neutrino energy for reactor experiments with baseline length $L=52.5~{\rm km}$ in the NH case. The solid (black) curve corresponds to the standard oscillation (SD), while the dashed (red) curve denotes the NSI polluted oscillation probability calculated by assuming only one nonzero NSI parameter $\varepsilon_{e\tau}=0.03$. In addition, the standard mixing parameters $\sin^2 \theta^{}_{12} = 0.307$, $\sin^2 \theta^{}_{13} = 0.0242$, $\sin^2 \theta^{}_{23} = 0.446$, $\delta = 0$, $\Delta m^2_{21} = 7.54 \times 10^{-5}~{\rm eV}^2$, and $\Delta m^2_{31} = 2.43\times 10^{-3}~{\rm eV}^2$ have been used. 
}
\end{center}
\end{figure}

For illustration, we show the survival probability $P(\bar{\nu}_e \to \bar{\nu}_e)$ with and without NSIs in Fig.~\ref{fig:Pee}. The NSI polluted survival probability deviates from the standard one especially at energies around 3~MeV. For $\varepsilon_{e\tau}=0.03$, the minimum of the  probability is slightly larger than in the standard case, indicating that the measured value of $\theta_{12}$ should be smaller than its true value. Such a feature can be understood from the oscillation probability in Eq.~\eqref{eq:Papp} that a negative $\varepsilon_\phi$ is expected for $\varepsilon^{}_{e\tau} = 0.03$ and $\varepsilon^{}_{e\mu} = \phi^{}_{e\tau} = \delta = 0$, reducing the amplitude of the $\Delta m^2_{21}$-driven oscillation. Consequently, one requires a smaller $\theta_{12}$ compared to the true input parameter in order to reduce the effective amplitude so as to gain a better fit. On the other hand, a positive $\varepsilon^{}_\delta$ enhances the amplitudes of the $\Delta m^2_{31}$ and $\Delta m^2_{32}$-driven oscillation modes, suggesting that a larger $\theta_{13}$ can be extracted from the experimental data. These observations are also confirmed later by our numerical analysis in Section~\ref{sub:NSIfit}.

If $\varepsilon^{}_{e\mu} = 0.03$ is assumed instead, the NSI-polluted survival probability becomes smaller than the standard one around the oscillation maximum associated with $\Delta m^2_{21}$, which can be understood from the opposite signs in front of $\varepsilon^{}_{e\mu}$ and $\varepsilon^{}_{e\tau}$ in Eq.~\eqref{eq:epsphi}. In an analogous way, one arrives at the conclusion that a larger value of $\theta^{}_{12}$ or $\theta^{}_{13}$ than its true value is expected from the fit to experimental data. However, at the probability level, there will be no visible differences between the NH and IH cases, which applies to the case with either $\varepsilon^{}_{e\mu} = 0.03$ or $\varepsilon^{}_{e\tau} = 0.03$. Finally, we remark that the oscillation frequencies are not affected by the NSIs, which can be observed from Eq.~\eqref{eq:Papp}. Therefore, the NSIs affect the determination of the neutrino mass hierarchy at a medium-baseline reactor experiment by modifying the oscillation amplitudes instead of the oscillation frequencies.

\section{numerical analysis} \label{sec:numerics}

We proceed to perform a numerical analysis of NSI effects on the measurements of neutrino parameters. To this end, we employ the GLoBES software~\cite{Huber:2004ka,Huber:2007ji} and consider the configuration of the JUNO experiment described in detail in Ref.~\cite{Li:2013zyd}. Explicitly, we take the reactor thermal power to be $P=35.8~{\rm GW}$ and the baseline length $L=52.5~{\rm km}$. The detector mass is assumed to be 20~kt together with the energy resolution $3\%/\sqrt{E}$ as a benchmark, where the energy is given in units of MeV. As discussed in Ref.~\cite{Li:2013zyd}, such an experimental setup allows one to determine the neutrino mass hierarchy at a confidence level of 4$\sigma$ for a six-year running (with 300 effective days per year). We further make use of a single overall factor $f$ together with the error $\delta f=3\%$ to parametrize the uncertainties of the total reactor antineutrino flux, the cross section of inverse beta decay, the fiducial mass, and the weight fraction of free protons.

The original Abstract Experiment Definition Language (AEDL) file for reactor antineutrino experiments in GLoBES \cite{Huber:2003pm} is modified for our purpose. The parametrization of reactor antineutrino flux is taken from Ref.~\cite{Murayama:2000iq}, in which the same fuel composition as in Ref.~\cite{Eguchi:2002dm} is assumed, and the cross section for the inverse beta decay $\bar{\nu}_e + p \to e^+ + n$ from Ref.~\cite{Vogel:1999zy} is adopted. The neutrino events are simulated by using the following true values for the oscillation parameters
\begin{eqnarray}\label{eq:paras}
\sin^2\theta^{\rm true}_{12} & = &  0.307 \pm 0.017 \; , \nonumber \\
\sin^2\theta^{\rm true}_{13} & = &  0.0242 \pm 0.0025 \; ,\nonumber \\
\sin^2\theta^{\rm true}_{23} & = &  0.446 \pm 0.007 \; ,\nonumber \\
\left(\Delta m^2_{21}\right)^{\rm true} & = & (7.54 \pm 0.24) \times 10^{-5}~{\rm eV}^2 \; , \nonumber \\
\left| \Delta m^2_{31} \right|^{\rm true} & = & (2.43 \pm 0.07) \times 10^{-3}~{\rm eV}^2 \; ,
\end{eqnarray}
together with the NSI parameters. The running time is taken to be five years, equivalent to the JUNO setup in Ref.~\cite{Li:2013zyd}. The total number of non-oscillated neutrino events is around $3.9\times 10^5$, and that of the oscillated ones is about $1.2\times 10^5$ for the central values of standard mixing parameters in Eq.~\eqref{eq:paras}. Note that no backgrounds are assumed in our simulations. The simulated data are then processed using the standard $\chi^2$ analysis through the following $\chi^2$ function
\begin{eqnarray}\label{eq:chi2}
\chi^2 = \min_p \sum^{}_i \frac{\left[N_i (p^{\rm true},\varepsilon^{\rm true})-N_i (p,\varepsilon=0)\right]^2}{N_i (p^{\rm true},\varepsilon^{\rm true})} + {\rm priors} \; ,
\end{eqnarray}
where $N_i$ denotes the number of events in the $i$-th energy bin, the parameter vector $p$ contains the standard oscillation parameters and the systematical errors, and $\varepsilon$ represents the NSI parameters. In the fit, all the standard oscillation parameters for the $\bar \nu_e \to \bar \nu_e$ channel are marginalized over, but the NSI parameters are fixed to zero. The reason is that we are interested in how the standard oscillation fit is modified when there are NSIs involved. The prior terms implement external input from other experiments and have the form $(p-p^{\rm true})^2/\sigma^2_p$ with $\sigma_p$ being the corresponding externally given uncertainty.

\subsection{Energy distribution}

\begin{figure}[t]
\begin{center}
\subfigure{%
\hspace{-1.4cm}
\includegraphics[width=0.64\textwidth]{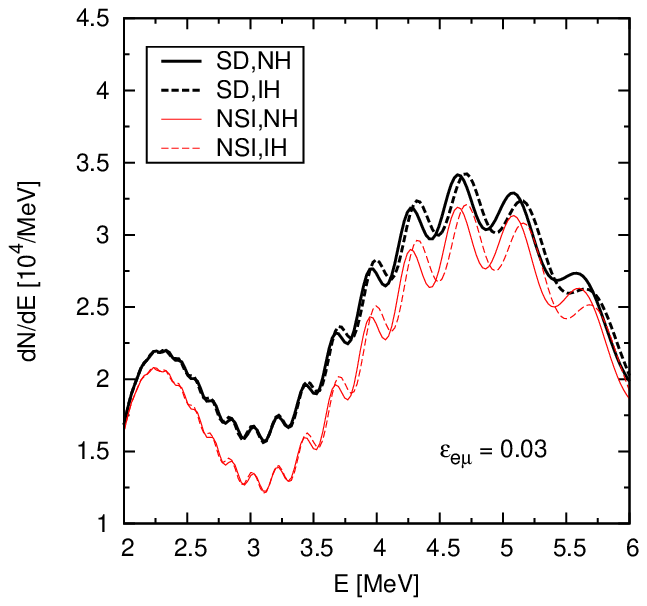}        }%
\subfigure{%
\hspace{-2.4cm}
\includegraphics[width=0.64\textwidth]{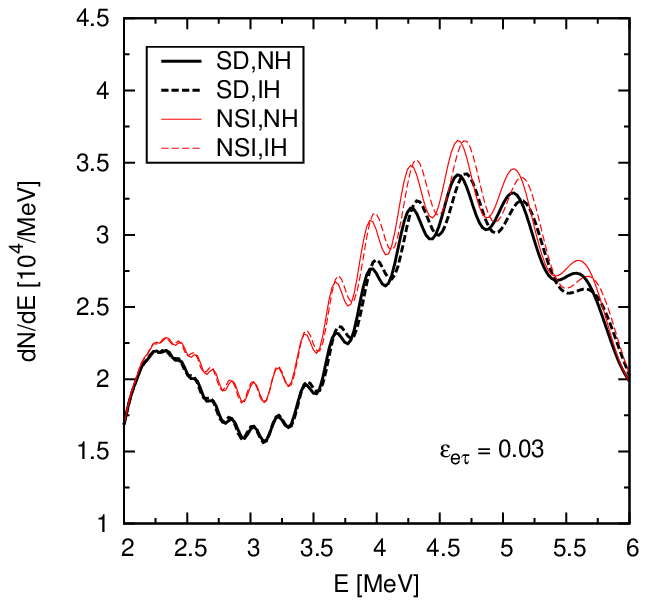}        }
\end{center}
\vspace{-0.5cm}
\caption{\label{fig:dNdE} The expected energy distribution of the antineutrinos with the energy resolution of $3\%/\sqrt{E}$. The thick (black) curves stand for the standard oscillation case, and the thin (red) curves for the NSI case. The chosen NSI parameters are given in each plot, while the solid and dashed curves correspond to the NH and IH cases, respectively. 
}
\end{figure}

First, in Fig.~\ref{fig:dNdE}, we show the expected energy distribution of antineutrino events for the JUNO setup. In each plot, the thick (black) curves stand for the distributions for the standard case, whereas the thin (red) curves correspond to the distributions when the NSI effects are included. Independently of the neutrino mass hierarchy, one can see a clear distinction between the black (thick) and red (thin) curves. However, such a spectrum shift can be absorbed into the mixing angle $\theta_{12}$, which does not lead to significant contributions to the $\chi^2$ function but may change the fitted value of $\sin^2\theta_{12}$ dramatically. When $\varepsilon_{e\mu}$ is positive, the event rate decreases compared to the standard case, implying that the extracted value of $\theta_{12}$ is slightly larger than its true value. Similarly, a nonvanishing $\varepsilon_{e\tau}$ would result in a smaller fitted value for $\theta_{12}$. In both cases, the amplitude of high-frequency oscillations is enhanced, so the a larger fitted value of $\theta^{}_{13}$ is expected. These observations are in accordance with our discussions on the survival probability in Section~\ref{sec:formalism}, and in agreement with our numerical fit in the following subsection. The impact of NSIs on the event rates and measurements of standard oscillation parameters has also been discussed in Ref.~\cite{Khan:2013hva}, where a different experimental setup was considered.

\subsection{NSI effects in the parameter fit}
\label{sub:NSIfit}
\begin{figure}[!h]
\begin{center}\vspace{-0.4cm}
\subfigure{%
\hspace{-1.4cm}
\includegraphics[width=0.64\textwidth]{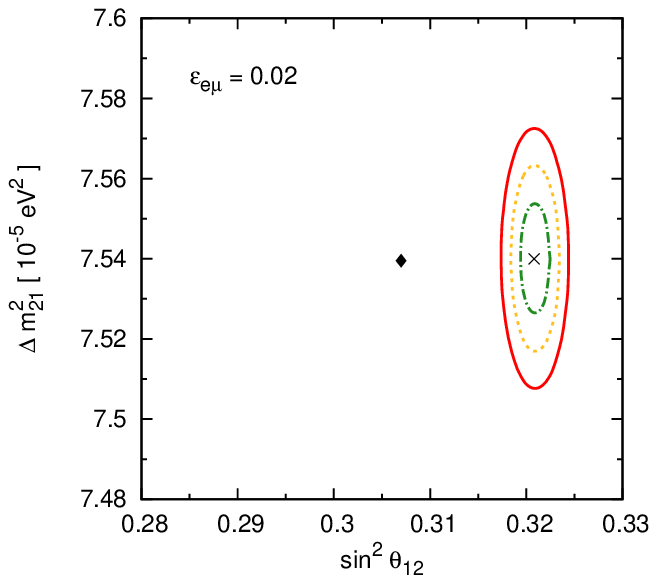}        }%
\subfigure{%
\hspace{-2.4cm}
\includegraphics[width=0.64\textwidth]{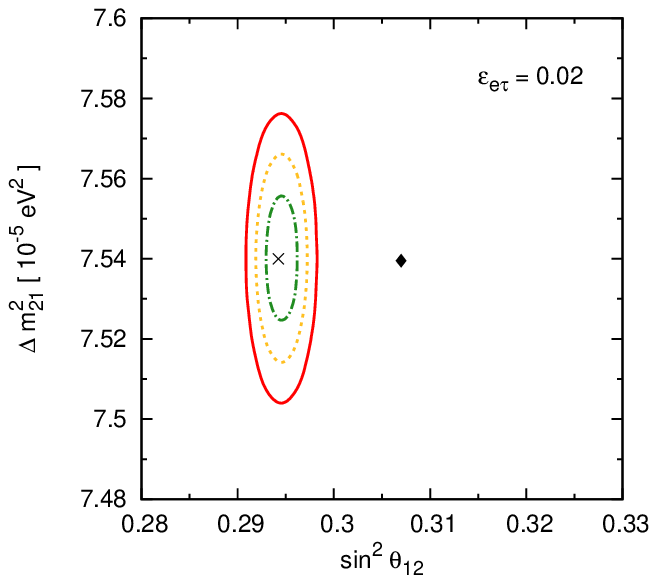}        }
\end{center}
\vspace{-0.8cm}
\caption{\label{fig:theta_mass1} The errors induced by NSIs in fitting $\theta_{12}$ and $\Delta m^2_{21}$ to the simulated data. The black diamonds indicate the true values of the neutrino parameters, whereas the crosses correspond to the extracted parameters. The dotted-dashed (green), dotted (yellow), and solid (red) curves stand for the 1$\sigma$, 2$\sigma$, and 3$\sigma$ C.L., respectively, while the input NSI parameters are given in each plot. 
}
\begin{center}\vspace{0.1cm}
\subfigure{%
\hspace{-1.4cm}
\includegraphics[width=0.64\textwidth]{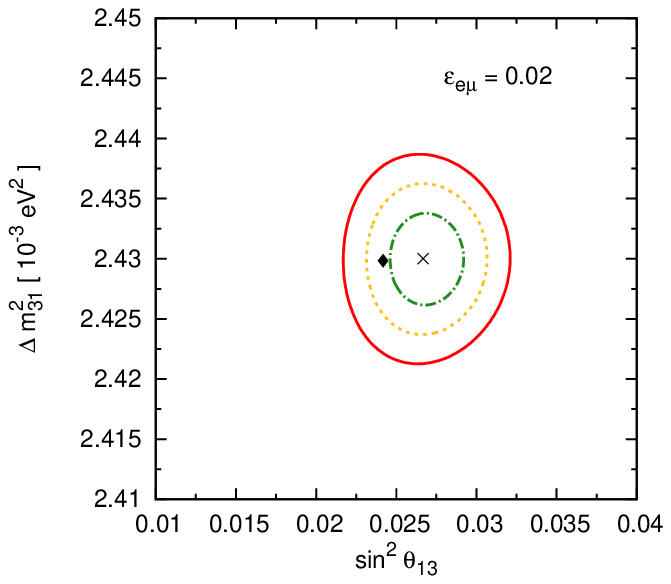}        }%
\subfigure{%
\hspace{-2.4cm}
\includegraphics[width=0.64\textwidth]{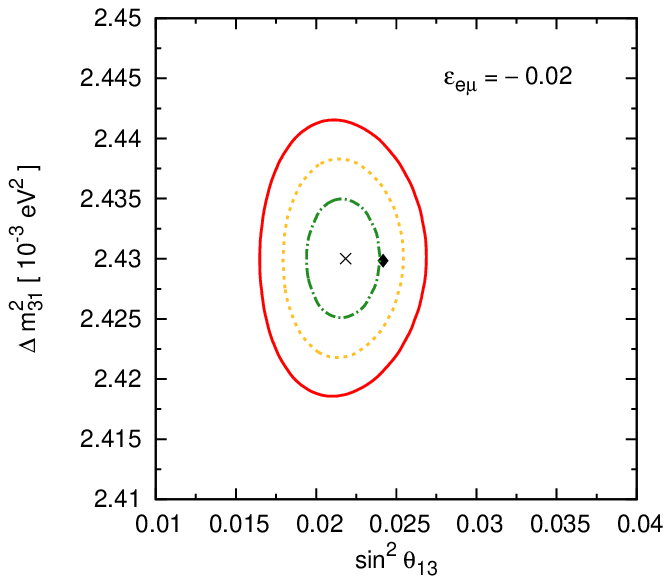}        }
\end{center}
\vspace{-0.8cm}
\caption{\label{fig:theta_mass2} The errors induced by NSIs in fitting $\theta_{13}$ and $\Delta m^2_{31}$ to the simulated data. The black diamonds indicate the true values of the neutrino parameters, whereas the crosses correspond to the extracted parameters. The dotted-dashed (green), dotted (yellow), and solid (red) curves stand for the 1$\sigma$, 2$\sigma$, and 3$\sigma$ C.L., respectively, while the input NSI parameters are given in each plot. 
}
\end{figure}

Second, in order to examine the NSI-induced offsets in the standard oscillation fit, we use the true parameters from Eq.~\eqref{eq:paras} together with a non-vanishing NSI parameter to generate the neutrino events. Then, in the fit, all the standard oscillation parameters are marginalized over, while the NSI parameters are fixed to zero. Assuming two degrees of freedom, we show the effects of NSIs in the fit in Figs.~\ref{fig:theta_mass1} and \ref{fig:theta_mass2} for NH and some benchmark NSI parameters.

Using Fig.~\ref{fig:theta_mass1}, one finds that the deviation of the best-fit $\sin^2\theta_{12}$ from its true value is remarkable. Even for a relatively small values of the NSI parameters, the true $\theta_{12}$ may be ruled out erroneously. Consistent with our analytical studies in the previous section, a positive $\varepsilon_{e\mu}$ leads to a fake $\theta_{12}$ larger than its true value, whereas a positive $\varepsilon_{e\tau}$ signifies a positive $\varepsilon_{\phi}$, and hence, a smaller fitted $\theta_{12}$ is favored.

As for $\theta_{13}$, one observes from Fig.~\ref{fig:theta_mass2} that the best-fit value only mildly deviates from its true value, which can be viewed as a result of the moderate sensitivity for the JUNO experiment to $\theta_{13}$. The result for $\varepsilon^{}_{e\tau} = 0.02$ is quite similar to that for $\varepsilon^{}_{e\mu} = 0.02$ in the left panel. For comparison, we show in the right panel that the best-fit value of $\theta^{}_{13}$ becomes smaller than the true value if $\varepsilon^{}_{e\mu} = -0.02$. As mentioned before, the determination of the neutrino mass-squared differences are essentially not spoiled by NSI effects, and both $\Delta m^2_{21}$ and $\Delta m^2_{31}$ can be fixed at a very-high confidence level as shown in the plots. It is worthwhile to note that the precisions of $\Delta m^2_{21}$ and $\Delta m^2_{31}$ are slightly better for larger values of $\theta^{}_{12}$ and $\theta^{}_{13}$, respectively, as observed from Figs.~\ref{fig:theta_mass1} and \ref{fig:theta_mass2}.

\subsection{Discovery reach}
\begin{figure}[!h]
\begin{center}\vspace{-0.4cm}
\subfigure{%
\hspace{-1.4cm}
\includegraphics[width=0.64\textwidth]{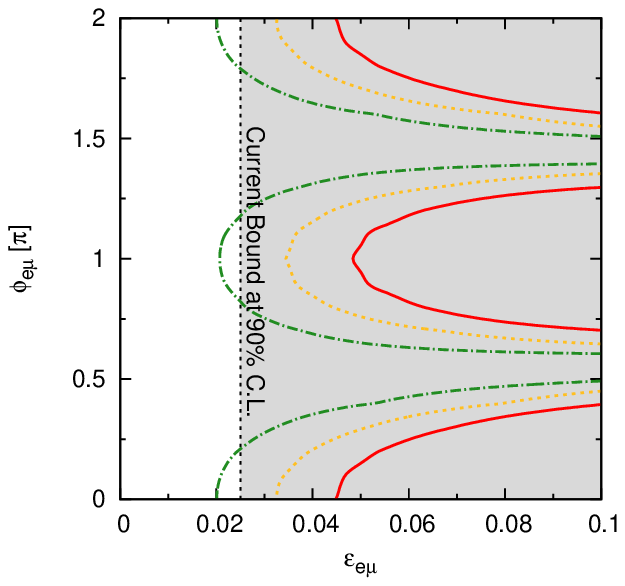}        }%
\subfigure{%
\hspace{-2.4cm}
\includegraphics[width=0.64\textwidth]{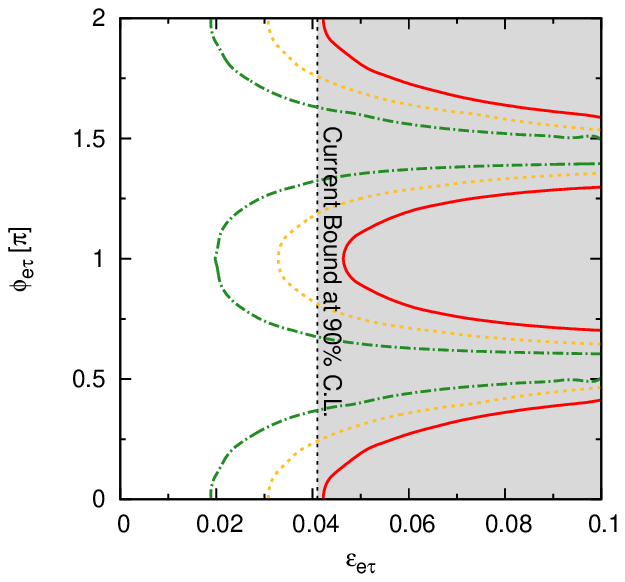}        }
\end{center}
\vspace{-0.8cm}
\caption{\label{fig:eps-phi} The discovery reach for $\varepsilon_{e\mu}$ and $\phi_{e\mu}$ (left plot), and for $\varepsilon_{e\tau}$ and $\phi_{e\tau}$ (right plot), in the NH case. Here the Dirac CP-violating phase $\delta = 0$ has been assumed, and the current 90\% C.L. constraints on the NSI parameters are represented by using gray shading. The dashed-dotted (green), dotted (yellow), and solid (red) curves stand for the 1$\sigma$, 2$\sigma$, and 3$\sigma$ C.L., respectively. 
}
\vspace{0.25cm}
\begin{center}\vspace{-0.1cm}
\subfigure{%
\hspace{-1.4cm}
\includegraphics[width=0.64\textwidth]{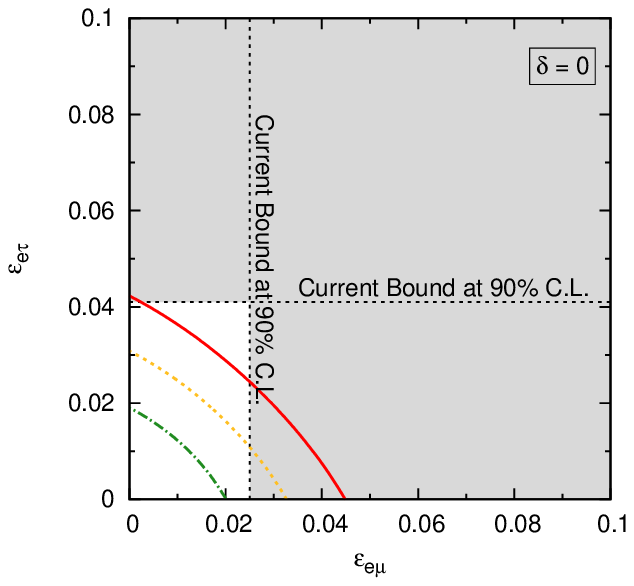}        }%
\subfigure{%
\hspace{-2.4cm}
\includegraphics[width=0.64\textwidth]{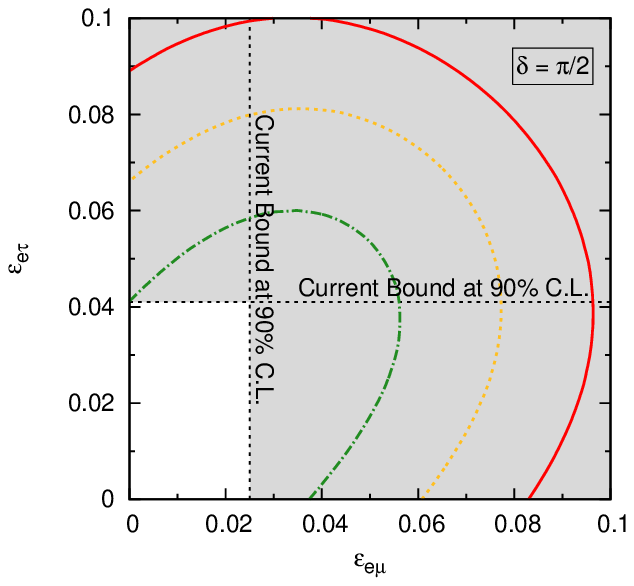}        }
\end{center}
\vspace{-0.8cm}
\caption{\label{fig:eps-eps} The discovery reach for the $\varepsilon^{}_{e\mu}$ and $\varepsilon^{}_{e\tau}$ in the case of $\delta = 0$ (left plot) and $\delta = \pi/2$ (right plot). Here the non-standard phases $\phi^{}_{e\mu}$ and $\phi^{}_{e\tau}$ are set to zero, and the current 90\% C.L. constraints on the NSI parameters are represented by using gray shading. The dashed-dotted (green), dotted (yellow), and solid (red) curves stand for the 1$\sigma$, 2$\sigma$, and 3$\sigma$ C.L., respectively. 
}
\end{figure}
Third, we continue to discuss the experimental prospect of detecting NSI effects. For this purpose, we define the discovery reach as the $\varepsilon$ ranges where the quality of a standard oscillation fit is below a given confidence level~\cite{Kopp:2007ne}. The numerical results are shown in Figs.~\ref{fig:eps-phi} and \ref{fig:eps-eps}.
In the left panel of Fig.~\ref{fig:eps-phi}, only the NSI parameters $\varepsilon^{}_{e\mu}$ and $\phi^{}_{e\mu}$ are switched on, while in the right panel only $\varepsilon^{}_{e\tau}$ and $\phi^{}_{e\tau}$. One can observe that the best sensitivity to $\varepsilon^{}_{e\mu}$ or $\varepsilon^{}_{e\tau}$ appears at $\phi \sim 0$ or $\pi$. This is in agreement with Eqs.~\eqref{eq:epsphi} and \eqref{eq:epsdelta}, in which the maximal values of $|\varepsilon_\phi|$ and $|\varepsilon_\delta|$ are reached for $\cos\phi^{}_{e\mu}=1$ or $\cos\phi^{}_{e\tau}=1$. It is worth mentioning that a nonzero value of the Dirac CP-violating phase $\delta$ results in only a global shift of the contour lines in Fig.~\ref{fig:eps-phi}. Current bounds on $\varepsilon^{}_{e\mu}$ and $\varepsilon^{}_{e\tau}$ at the $90\%$ C.L., i.e., $\varepsilon^{}_{e\mu} < 0.025$ and $\varepsilon^{}_{e\tau} < 0.041$, are already very stringent~\cite{Biggio:2009nt}. The medium-baseline reactor experiments can improve the bound on $\varepsilon^{}_{e\tau}$ to below $0.03$, but they can hardly set any constraints on $\varepsilon^{}_{e\mu}$.

In Fig.~\ref{fig:eps-eps}, we switch on both $\varepsilon^{}_{e\mu}$ and $\varepsilon^{}_{e\tau}$, but set the corresponding phases to zero. In this case, the input value of $\delta$ changes the sensitivity dramatically. As shown in the right panel of Fig.~\ref{fig:eps-eps}, a maximal CP-violating phase $\delta=\pi/2$ leads to a vanishing value of $\varepsilon_\delta$ and a suppressed value of $\varepsilon_\phi$ due to the approximate $\mu$-$\tau$ symmetry. Compared to the current bounds on the NSI parameters, a medium-baseline reactor experiment only moderately improves the constraints for $\delta = 0$. In this case, a $3\sigma$ hint can be obtained if both $\varepsilon^{}_{e\mu}$ and $\varepsilon^{}_{e\tau}$ are larger than 0.02. We remark that the discovery reach on NSIs depends strongly on the assumed uncertainty of $\theta_{13}$. In our numerical analysis, a relatively optimistic uncertainty on $\theta_{13}$ is assumed, which actually relies on the precision of the measurement of $\theta_{13}$ in ongoing and future non-reactor experiments (e.g., the long-baseline accelerator experiments T2K and NO$\nu$A). In general, NSIs induce conflicts between measurement obtained in different type of experiments, and therefore, a combined analysis of both short-, medium-baseline reactor experiments, and accelerator experiments will be more powerful in constraining the NSI parameters. However, such an analysis is beyond the scope of this work and will be performed elsewhere.

Note that we have concentrated on NSIs in the production and detection of reactor antineutrinos, and ignored NSIs in the propagation, which could also be important for solar~\cite{Miranda:2004nb,Bolanos:2008km,Escrihuela:2009up}, atmospheric~\cite{Fornengo:2001pm,Huber:2001zw,Mitsuka:2011ty,Escrihuela:2011cf,Ohlsson:2013epa,Esmaili:2013fva}, and long-baseline neutrino oscillation experiments~\cite{Blennow:2007pu,Ribeiro:2007ud,Blennow:2008eb,Adamson:2013ovz,Kopp:2007mi}. Source and detector NSIs have been examined in Ref.~\cite{Leitner:2011aa} for the Daya Bay experiment, where it has been found that the effective mixing parameters $\sin^2 2\tilde\theta_{13}$ and $\Delta \tilde m^2_{32}$ are shifted from 0.1 to 0.105 and $2.45 \times 10^{-3} \, {\rm eV}^2$ to $2.2 \times 10^{-3} \, {\rm eV}^2$, respectively, if $\varepsilon_{e\mu} = \varepsilon_{e\tau} = 0.02$ is assumed. In Ref.~\cite{Khan:2013hva}, NSI effects at the short- and medium-baseline reactor experiments are studied. However, no experimental sensitivities to the NSI parameters $\varepsilon_{e\mu}$ and $\varepsilon_{e\tau}$ have been presented. To our knowledge, source and detector NSIs in reactor antineutrino experiments have previously been discussed only in Refs.~\cite{Ohlsson:2008gx,Leitner:2011aa,Kopp:2007ne,Khan:2013hva,Escrihuela:2009up}.

\subsection{Impact on the neutrino mass hierarchy determination}

Fourth, the NSI parameters modify the oscillation probability, and hence may diminish or improve the experimental sensitivity to the neutrino mass hierarchy. In order to see the NSI effects in the neutrino mass hierarchy fit, we simulate the data in the NH case with nonzero NSI parameters, and then perform a standard oscillation fit with either NH or IH.

\begin{figure}[t]
\begin{center}
\subfigure{%
\hspace{-1.5cm}
\includegraphics[width=0.64\textwidth]{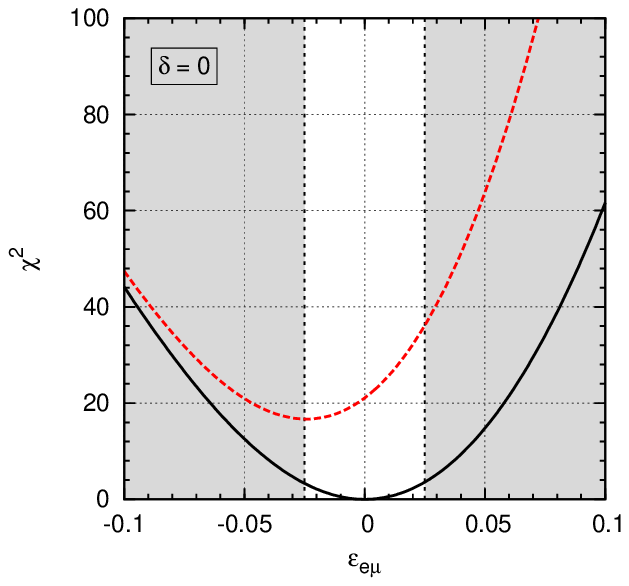}        }%
\subfigure{%
\hspace{-2.5cm}
\includegraphics[width=0.64\textwidth]{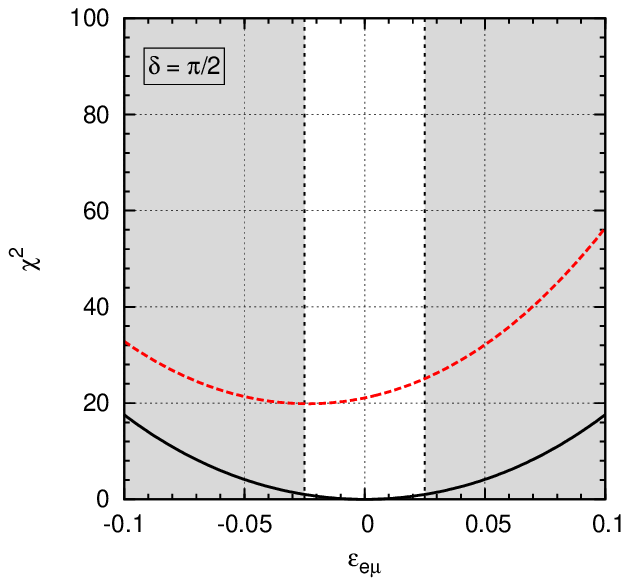}        } \\
\vspace{-0.2cm}
\subfigure{%
\hspace{-1.5cm}
\includegraphics[width=0.64\textwidth]{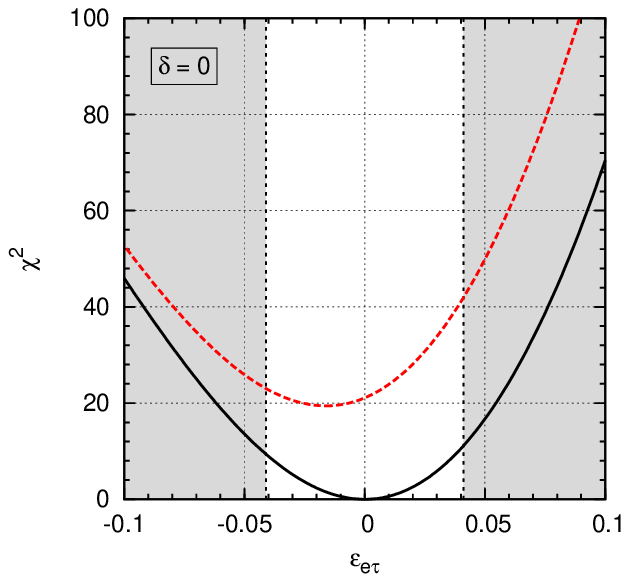}        }%
\subfigure{%
\hspace{-2.5cm}
\includegraphics[width=0.64\textwidth]{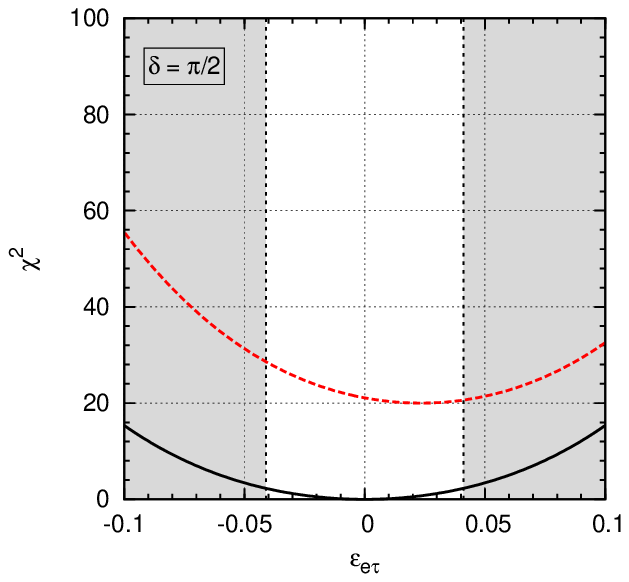}        }
\end{center}
\vspace{-0.8cm}
\caption{\label{fig:chi2} The $\chi^2$ for the mass hierarchy discrimination shown as a function of the NSI parameters. The events are generated by assuming NH, while the solid (black) and dashed (red) curves correspond to the standard-oscillation fit with NH and IH, respectively. 
}
\end{figure}

The dependence of the minimal $\chi^2$ on the NSI parameters is shown in Fig.~\ref{fig:chi2}, where one can observe that the $\chi^2$ can be slightly diminished compared to the standard case for specific choices of the NSI parameters. As in the upper-left plot, the $\chi^2$ value is reduced approximately from 20 to 16 for $\varepsilon_{e\mu} \simeq -0.025$ and $\delta=0$, and the difference between the NH and IH fit decreases to 12. On the other hand, a positive $\varepsilon_{e\mu}$ or $\varepsilon_{e\tau}$ leads to a larger $\chi^2$ in the wrong hierarchy fit. The deterioration effects in the neutrino mass hierarchy fit are mainly due to the mimicking effects on $\theta_{13}$. Namely, for some choices of NSI parameters, the probability difference between the NH and IH cases defined in Eq.~\eqref{eq:deltaP} can be smaller than the standard situation, and thus the $\chi^2$ fit becomes a little worse.

However, if current bounds on the NSI parameters are taken into account, the impact of NSIs on the determination of neutrino mass hierarchy is not important. As shown in the right column of Fig.~\ref{fig:chi2}, if $\delta = \pi/2$ is assumed, the NSI effects will be even smaller. In a different way, one can generate the neutrino events by assuming IH, and fit the data with either NH or IH. However, we have confirmed that the difference between these two approaches is insignificant.

Compared to the $\chi^2$ analysis of the sensitivity to the neutrino mass hierarchy at JUNO in Ref.~\cite{Li:2013zyd}, our minimal $\chi^2$ for vanishing NSI parameters for the wrong hierarchy fit $\chi^2 \sim 20$ is slightly larger. This difference can be ascribed to our optimistic treatment of the systematic uncertainties and the backgrounds, and to a larger number of simulated neutrino events.

\section{Summary and Conclusions}
\label{sec:summary}

A reactor antineutrino experiment with a medium-baseline of around 50~km has been proposed to pin down the neutrino mass hierarchy, and to precisely measure leptonic mixing parameters and neutrino mass-squared differences. With the expected high-precision measurements of neutrino parameters in this kind of experiment, one can probe new physics beyond the standard-oscillation paradigm as sub-leading effects in neutrino flavor conversion.

In this Letter, we have investigated the impact of NSIs on the medium-baseline reactor antineutrino experiments. For reactor antineutrinos, only the NSI parameters $\varepsilon^{}_{e\mu}$ and $\varepsilon^{}_{e\tau}$, and the corresponding phases $\phi^{}_{e\mu}$ and $\phi^{}_{e\tau}$, are relevant. First of all, we demonstrate that the true value of the leptonic mixing angle $\theta^{}_{12}$ can be erroneously ruled out at more than $3\sigma$ C.L.~if the NSI parameter $\varepsilon^{}_{e\mu}$ or $\varepsilon^{}_{e\tau}$ is as large as 0.02. However, the extraction of the leptonic mixing angle $\theta^{}_{13}$ from the experimental data is rarely affected, since the experimental sensitivity to $\theta^{}_{13}$ is not as high as the one to $\theta^{}_{12}$. Then, we show the discovery reach of NSI effects at the medium-baseline experiment. It turns out that the CP-violating phases play an important role in constraining $\varepsilon^{}_{e\mu}$ and $\varepsilon^{}_{e\tau}$. In the most optimistic case, a $3\sigma$ hint for NSI effects can be obtained when both $\varepsilon^{}_{e\mu}$ and $\varepsilon^{}_{e\tau}$ are larger than 0.02. Finally, the impact of NSIs on the determination of the neutrino mass hierarchy is considered. In principle, NSIs can diminish or enhance significantly the experimental power in discriminating neutrino mass hierarchies. However, when the current bounds on the NSI parameters are taken into account, such effects become insignificant.

It is worthwhile to emphasize that the precision measurements of neutrino parameters in the ongoing and forthcoming neutrino oscillation experiments provide a good opportunity to probe NSIs and the underlying new physics, which is complementary to the direct searches at collider experiments. A complete analysis of the NSI effects should combine all relevant experiments, and obviously deserves further detailed studies.

\begin{acknowledgments}
H.Z. would like to thank the G{\"o}ran Gustafsson Foundation for financial support, and the KTH Royal Institute of Technology for warm hospitality, where part of this work was performed. The authors are indebted to Joachim Kopp for useful discussions. This work was supported by the Swedish Research Council (Vetenskapsr{\aa}det), contract no. 621-2011-3985 (T.O.) and the Max Planck Society through the Strategic Innovation Fund in the project MANITOP (H.Z.).
\end{acknowledgments}




\begin{thebibliography}{24}
\expandafter\ifx\csname natexlab\endcsname\relax\def\natexlab#1{#1}\fi
\expandafter\ifx\csname bibnamefont\endcsname\relax
  \def\bibnamefont#1{#1}\fi
\expandafter\ifx\csname bibfnamefont\endcsname\relax
  \def\bibfnamefont#1{#1}\fi
\expandafter\ifx\csname citenamefont\endcsname\relax
  \def\citenamefont#1{#1}\fi
\expandafter\ifx\csname url\endcsname\relax
  \def\url#1{\texttt{#1}}\fi
\expandafter\ifx\csname urlprefix\endcsname\relax\def\urlprefix{URL }\fi
\providecommand{\bibinfo}[2]{#2}
\providecommand{\eprint}[2][]{\url{#2}}

\bibitem[{\citenamefont{An et~al.}(2012)}]{An:2012eh}
\bibinfo{author}{\bibfnamefont{F.}~\bibnamefont{An}} \bibnamefont{et~al.}
  (\bibinfo{collaboration}{DAYA-BAY Collaboration}),
  \bibinfo{journal}{Phys.~Rev.~Lett.} \textbf{\bibinfo{volume}{108}},
  \bibinfo{pages}{171803} (\bibinfo{year}{2012}), \eprint{1203.1669}.

\bibitem[{\citenamefont{Ahn et~al.}(2012)}]{Ahn:2012nd}
\bibinfo{author}{\bibfnamefont{J.}~\bibnamefont{Ahn}} \bibnamefont{et~al.}
  (\bibinfo{collaboration}{RENO collaboration}),
  \bibinfo{journal}{Phys.~Rev.~Lett.} \textbf{\bibinfo{volume}{108}},
  \bibinfo{pages}{191802} (\bibinfo{year}{2012}), \eprint{1204.0626}.

\bibitem[{\citenamefont{Abe et~al.}(2012)}]{Abe:2012tg}
\bibinfo{author}{\bibfnamefont{Y.}~\bibnamefont{Abe}} \bibnamefont{et~al.}
  (\bibinfo{collaboration}{Double Chooz Collaboration}),
  \bibinfo{journal}{Phys.~Rev.} \textbf{\bibinfo{volume}{D86}},
  \bibinfo{pages}{052008} (\bibinfo{year}{2012}), \eprint{1207.6632}.

\bibitem[{\citenamefont{An et~al.}(2013)}]{An:2012bu}
\bibinfo{author}{\bibfnamefont{F.}~\bibnamefont{An}} \bibnamefont{et~al.}
  (\bibinfo{collaboration}{Daya Bay Collaboration}),
  \bibinfo{journal}{Chin.~Phys.} \textbf{\bibinfo{volume}{C37}},
  \bibinfo{pages}{011001} (\bibinfo{year}{2013}), \eprint{1210.6327}.

\bibitem[{\citenamefont{Abe et~al.}(2013)}]{Abe:2013xua}
\bibinfo{author}{\bibfnamefont{K.}~\bibnamefont{Abe}} \bibnamefont{et~al.}
  (\bibinfo{collaboration}{T2K Collaboration}), \bibinfo{journal}{Phys.~Rev.}
  \textbf{\bibinfo{volume}{D88}}, \bibinfo{pages}{032002}
  (\bibinfo{year}{2013}), \eprint{1304.0841}.

\bibitem[{\citenamefont{Kettell et~al.}(2013)\citenamefont{Kettell, Ling, Qian,
  Yeh, Zhang et~al.}}]{Kettell:2013eos}
\bibinfo{author}{\bibfnamefont{S.}~\bibnamefont{Kettell}},
  \bibinfo{author}{\bibfnamefont{J.}~\bibnamefont{Ling}},
  \bibinfo{author}{\bibfnamefont{X.}~\bibnamefont{Qian}},
  \bibinfo{author}{\bibfnamefont{M.}~\bibnamefont{Yeh}},
  \bibinfo{author}{\bibfnamefont{C.}~\bibnamefont{Zhang}}, \bibnamefont{et~al.}
  (\bibinfo{year}{2013}), \eprint{1307.7419}.

\bibitem[{\citenamefont{Li et~al.}(2013)\citenamefont{Li, Cao, Wang, and
  Zhan}}]{Li:2013zyd}
\bibinfo{author}{\bibfnamefont{Y.-F.} \bibnamefont{Li}},
  \bibinfo{author}{\bibfnamefont{J.}~\bibnamefont{Cao}},
  \bibinfo{author}{\bibfnamefont{Y.}~\bibnamefont{Wang}}, \bibnamefont{and}
  \bibinfo{author}{\bibfnamefont{L.}~\bibnamefont{Zhan}},
  \bibinfo{journal}{Phys.~Rev.} \textbf{\bibinfo{volume}{D88}},
  \bibinfo{pages}{013008} (\bibinfo{year}{2013}), \eprint{1303.6733}.

\bibitem[{\citenamefont{Seo}(2013)}]{RENO50}
\bibinfo{author}{\bibfnamefont{S.~H.} \bibnamefont{Seo}}
  (\bibinfo{year}{2013}), \bibinfo{note}{talk given at the {\it International
  Workshop on ``RENO-50" toward Neutrino Mass Hierarchy}, 13-14 June 2013,
  Seoul National University, Korea}.

\bibitem[{\citenamefont{Wolfenstein}(1978)}]{Wolfenstein:1977ue}
\bibinfo{author}{\bibfnamefont{L.}~\bibnamefont{Wolfenstein}},
  \bibinfo{journal}{Phys.~Rev.} \textbf{\bibinfo{volume}{D17}},
  \bibinfo{pages}{2369} (\bibinfo{year}{1978}).

\bibitem[{\citenamefont{Valle}(1987)}]{Valle:1987gv}
\bibinfo{author}{\bibfnamefont{J.~W.~F.}~\bibnamefont{Valle}},
  \bibinfo{journal}{Phys.~Lett.} \textbf{\bibinfo{volume}{B199}},
  \bibinfo{pages}{432} (\bibinfo{year}{1987}).

\bibitem[{\citenamefont{Guzzo et~al.}(1991)\citenamefont{Guzzo,
  Masiero, and Petcov}}]{Guzzo:1991hi}
\bibinfo{author}{\bibfnamefont{M.}~\bibnamefont{Guzzo}},
  \bibinfo{author}{\bibfnamefont{A.}~\bibnamefont{Masiero}},
  \bibinfo{author}{\bibfnamefont{S.}~\bibnamefont{Petcov}},
  \bibinfo{journal}{Phys.~Lett.} \textbf{\bibinfo{volume}{B260}},
  \bibinfo{pages}{154} (\bibinfo{year}{1991}).

\bibitem[{\citenamefont{Roulet}(1991)}]{Roulet:1991sm}
\bibinfo{author}{\bibfnamefont{E.}~\bibnamefont{Roulet}},
  \bibinfo{journal}{Phys.~Rev.} \textbf{\bibinfo{volume}{D44}},
  \bibinfo{pages}{935} (\bibinfo{year}{1991}).

\bibitem[{\citenamefont{Grossman}(1995)}]{Grossman:1995wx}
\bibinfo{author}{\bibfnamefont{Y.}~\bibnamefont{Grossman}},
  \bibinfo{journal}{Phys.~Lett.} \textbf{\bibinfo{volume}{359}},
  \bibinfo{pages}{141} (\bibinfo{year}{1995}), \eprint{hep-ph/9507344}.

\bibitem[{\citenamefont{Ohlsson}(2013)}]{Ohlsson:2012kf}
\bibinfo{author}{\bibfnamefont{T.}~\bibnamefont{Ohlsson}},
  \bibinfo{journal}{Rept.~Prog.~Phys.} \textbf{\bibinfo{volume}{76}},
  \bibinfo{pages}{044201} (\bibinfo{year}{2013}), \eprint{1209.2710}.

\bibitem[{\citenamefont{Ohlsson and Zhang}(2009)}]{Ohlsson:2008gx}
\bibinfo{author}{\bibfnamefont{T.}~\bibnamefont{Ohlsson}} \bibnamefont{and}
  \bibinfo{author}{\bibfnamefont{H.}~\bibnamefont{Zhang}},
  \bibinfo{journal}{Phys. Lett.} \textbf{\bibinfo{volume}{B671}},
  \bibinfo{pages}{99} (\bibinfo{year}{2009}), \eprint{0809.4835}.

\bibitem[{\citenamefont{Leitner et~al.}(2011)\citenamefont{Leitner,
  Malinsk{\'y}, Roskovec, and Zhang}}]{Leitner:2011aa}
\bibinfo{author}{\bibfnamefont{R.}~\bibnamefont{Leitner}},
  \bibinfo{author}{\bibfnamefont{M.}~\bibnamefont{Malinsk{\'y}}},
  \bibinfo{author}{\bibfnamefont{B.}~\bibnamefont{Roskovec}}, \bibnamefont{and}
  \bibinfo{author}{\bibfnamefont{H.}~\bibnamefont{Zhang}},
  \bibinfo{journal}{JHEP} \textbf{\bibinfo{volume}{1112}}, \bibinfo{pages}{001}
  (\bibinfo{year}{2011}), \eprint{1105.5580}.

\bibitem[{\citenamefont{Kopp et~al.}(2008)\citenamefont{Kopp, Lindner, Ota, and
  Sato}}]{Kopp:2007ne}
\bibinfo{author}{\bibfnamefont{J.}~\bibnamefont{Kopp}},
  \bibinfo{author}{\bibfnamefont{M.}~\bibnamefont{Lindner}},
  \bibinfo{author}{\bibfnamefont{T.}~\bibnamefont{Ota}}, \bibnamefont{and}
  \bibinfo{author}{\bibfnamefont{J.}~\bibnamefont{Sato}},
  \bibinfo{journal}{Phys.~Rev.} \textbf{\bibinfo{volume}{D77}},
  \bibinfo{pages}{013007} (\bibinfo{year}{2008}), \eprint{0708.0152}.

\bibitem[{\citenamefont{Adhikari et~al.}(2012)\citenamefont{Adhikari,
  Chakraborty, Dasgupta, and Roy}}]{Adhikari:2012vc}
\bibinfo{author}{\bibfnamefont{R.}~\bibnamefont{Adhikari}},
  \bibinfo{author}{\bibfnamefont{S.}~\bibnamefont{Chakraborty}},
  \bibinfo{author}{\bibfnamefont{A.}~\bibnamefont{Dasgupta}}, \bibnamefont{and}
  \bibinfo{author}{\bibfnamefont{S.}~\bibnamefont{Roy}},
  \bibinfo{journal}{Phys.~Rev.} \textbf{\bibinfo{volume}{D86}},
  \bibinfo{pages}{073010} (\bibinfo{year}{2012}), \eprint{1201.3047}.

\bibitem[{\citenamefont{Khan et~al.}(2013)\citenamefont{Khan, McKay, and
  Tahir}}]{Khan:2013hva}
\bibinfo{author}{\bibfnamefont{A.~N.} \bibnamefont{Khan}},
  \bibinfo{author}{\bibfnamefont{D.~W.} \bibnamefont{McKay}}, \bibnamefont{and}
  \bibinfo{author}{\bibfnamefont{F.} \bibnamefont{Tahir}}
  (\bibinfo{year}{2013}), \eprint{1305.4350}.

\bibitem[{\citenamefont{Biggio et~al.}(2009)\citenamefont{Biggio, Blennow, and
  Fernandez-Martinez}}]{Biggio:2009nt}
\bibinfo{author}{\bibfnamefont{C.}~\bibnamefont{Biggio}},
  \bibinfo{author}{\bibfnamefont{M.}~\bibnamefont{Blennow}}, \bibnamefont{and}
  \bibinfo{author}{\bibfnamefont{E.}~\bibnamefont{Fernandez-Martinez}},
  \bibinfo{journal}{JHEP} \textbf{\bibinfo{volume}{0908}}, \bibinfo{pages}{090}
  (\bibinfo{year}{2009}), \eprint{0907.0097}.

\bibitem[{\citenamefont{Antusch et~al.}(2006)\citenamefont{Antusch, Biggio,
  Fernandez-Martinez, Gavela, and L{\'o}pez-Pav{\'o}n}}]{Antusch:2006vwa}
\bibinfo{author}{\bibfnamefont{S.}~\bibnamefont{Antusch}},
  \bibinfo{author}{\bibfnamefont{C.}~\bibnamefont{Biggio}},
  \bibinfo{author}{\bibfnamefont{E.}~\bibnamefont{Fernandez-Martinez}},
  \bibinfo{author}{\bibfnamefont{M.~B.} \bibnamefont{Gavela}},
  \bibnamefont{and}
  \bibinfo{author}{\bibfnamefont{J.}~\bibnamefont{L{\'o}pez-Pav{\'o}n}},
  \bibinfo{journal}{JHEP} \textbf{\bibinfo{volume}{0610}}, \bibinfo{pages}{084}
  (\bibinfo{year}{2006}), \eprint{hep-ph/0607020}.

\bibitem[{\citenamefont{Schechter and Valle}(1980)}]{Schechter:1980gr}
\bibinfo{author}{\bibfnamefont{J.} \bibnamefont{Schechter}} \bibnamefont{and}
  \bibinfo{author}{\bibfnamefont{J.~W.~F.} \bibnamefont{Valle}},
  \bibinfo{journal}{Phys.~Rev.} \textbf{\bibinfo{volume}{D22}}, \bibinfo{pages}{2227}
  (\bibinfo{year}{1980}).

\bibitem[{\citenamefont{Mohapatra and Valle}(1986)}]{Mohapatra:1986bd}
\bibinfo{author}{\bibfnamefont{R.~N.} \bibnamefont{Mohapatra}} \bibnamefont{and}
  \bibinfo{author}{\bibfnamefont{J.~W.~F.} \bibnamefont{Valle}},
  \bibinfo{journal}{Phys.~Rev.} \textbf{\bibinfo{volume}{D34}}, \bibinfo{pages}{1642}
  (\bibinfo{year}{1986}).

\bibitem[{\citenamefont{Pilaftsis}(1991)}]{Pilaftsis:1991ug}
\bibinfo{author}{\bibfnamefont{A.} \bibnamefont{Pilaftsis}},
  \bibinfo{journal}{Z.~Phys.} \textbf{\bibinfo{volume}{C55}}, \bibinfo{pages}{275}
  (\bibinfo{year}{1992}), \eprint{hep-ph/9901206}.

\bibitem[{\citenamefont{Meloni et~al.}(2010)\citenamefont{Meloni, Ohlsson,
  Winter, and Zhang}}]{Meloni:2009cg}
\bibinfo{author}{\bibfnamefont{D.} \bibnamefont{Meloni}},
  \bibinfo{author}{\bibfnamefont{T.}~\bibnamefont{Ohlsson}},
  \bibinfo{author}{\bibfnamefont{W.}~\bibnamefont{Winter}}, \bibnamefont{and}
  \bibinfo{author}{\bibfnamefont{H.}~\bibnamefont{Zhang}},
  \bibinfo{journal}{JHEP} \textbf{\bibinfo{volume}{1004}},
  \bibinfo{pages}{041} (\bibinfo{year}{2010}), \eprint{0912.2735}.

\bibitem[{\citenamefont{Akhmedov et~al.}(2001)\citenamefont{Akhmedov, Huber,
  Lindner, and Ohlsson}}]{Akhmedov:2001kd}
\bibinfo{author}{\bibfnamefont{E.~K.} \bibnamefont{Akhmedov}},
  \bibinfo{author}{\bibfnamefont{P.}~\bibnamefont{Huber}},
  \bibinfo{author}{\bibfnamefont{M.}~\bibnamefont{Lindner}}, \bibnamefont{and}
  \bibinfo{author}{\bibfnamefont{T.}~\bibnamefont{Ohlsson}},
  \bibinfo{journal}{Nucl.~Phys.} \textbf{\bibinfo{volume}{B608}},
  \bibinfo{pages}{394} (\bibinfo{year}{2001}), \eprint{hep-ph/0105029}.

\bibitem[{\citenamefont{Jacobson and Ohlsson}(2004)}]{Jacobson:2003wc}
\bibinfo{author}{\bibfnamefont{M.}~\bibnamefont{Jacobson}} \bibnamefont{and}
  \bibinfo{author}{\bibfnamefont{T.}~\bibnamefont{Ohlsson}},
  \bibinfo{journal}{Phys.~Rev.} \textbf{\bibinfo{volume}{D69}},
  \bibinfo{pages}{013003} (\bibinfo{year}{2004}), \eprint{hep-ph/0305064}.

\bibitem[{\citenamefont{Huber et~al.}(2005)\citenamefont{Huber, Lindner, and
  Winter}}]{Huber:2004ka}
\bibinfo{author}{\bibfnamefont{P.}~\bibnamefont{Huber}},
  \bibinfo{author}{\bibfnamefont{M.}~\bibnamefont{Lindner}}, \bibnamefont{and}
  \bibinfo{author}{\bibfnamefont{W.}~\bibnamefont{Winter}},
  \bibinfo{journal}{Comput.~Phys.~Commun.} \textbf{\bibinfo{volume}{167}},
  \bibinfo{pages}{195} (\bibinfo{year}{2005}), \bibinfo{note}{{\tt
  http://www.mpi-hd.mpg.de/lin/globes/}}, \eprint{hep-ph/0407333}.

\bibitem[{\citenamefont{Huber et~al.}(2007)\citenamefont{Huber, Kopp, Lindner,
  Rolinec, and Winter}}]{Huber:2007ji}
\bibinfo{author}{\bibfnamefont{P.}~\bibnamefont{Huber}},
  \bibinfo{author}{\bibfnamefont{J.}~\bibnamefont{Kopp}},
  \bibinfo{author}{\bibfnamefont{M.}~\bibnamefont{Lindner}},
  \bibinfo{author}{\bibfnamefont{M.}~\bibnamefont{Rolinec}}, \bibnamefont{and}
  \bibinfo{author}{\bibfnamefont{W.}~\bibnamefont{Winter}},
  \bibinfo{journal}{Comput.~Phys.~Commun.} \textbf{\bibinfo{volume}{177}},
  \bibinfo{pages}{432} (\bibinfo{year}{2007}), \eprint{hep-ph/0701187}.

\bibitem[{\citenamefont{Huber et~al.}(2003)\citenamefont{Huber, Lindner,
  Schwetz, and Winter}}]{Huber:2003pm}
\bibinfo{author}{\bibfnamefont{P.}~\bibnamefont{Huber}},
  \bibinfo{author}{\bibfnamefont{M.}~\bibnamefont{Lindner}},
  \bibinfo{author}{\bibfnamefont{T.}~\bibnamefont{Schwetz}}, \bibnamefont{and}
  \bibinfo{author}{\bibfnamefont{W.}~\bibnamefont{Winter}},
  \bibinfo{journal}{Nucl.~Phys.} \textbf{\bibinfo{volume}{B665}},
  \bibinfo{pages}{487} (\bibinfo{year}{2003}), \eprint{hep-ph/0303232}.

\bibitem[{\citenamefont{Murayama and Pierce}(2002)}]{Murayama:2000iq}
\bibinfo{author}{\bibfnamefont{H.}~\bibnamefont{Murayama}} \bibnamefont{and}
  \bibinfo{author}{\bibfnamefont{A.}~\bibnamefont{Pierce}},
  \bibinfo{journal}{Phys.~Rev.} \textbf{\bibinfo{volume}{D65}},
  \bibinfo{pages}{013012} (\bibinfo{year}{2002}), \eprint{hep-ph/0012075}.

\bibitem[{\citenamefont{Eguchi et~al.}(2003)}]{Eguchi:2002dm}
\bibinfo{author}{\bibfnamefont{K.}~\bibnamefont{Eguchi}} \bibnamefont{et~al.}
  (\bibinfo{collaboration}{KamLAND Collaboration}),
  \bibinfo{journal}{Phys.~Rev.~Lett.} \textbf{\bibinfo{volume}{90}},
  \bibinfo{pages}{021802} (\bibinfo{year}{2003}), \eprint{hep-ex/0212021}.

\bibitem[{\citenamefont{Vogel and Beacom}(1999)}]{Vogel:1999zy}
\bibinfo{author}{\bibfnamefont{P.}~\bibnamefont{Vogel}} \bibnamefont{and}
  \bibinfo{author}{\bibfnamefont{J.~F.} \bibnamefont{Beacom}},
  \bibinfo{journal}{Phys.~Rev.} \textbf{\bibinfo{volume}{D60}},
  \bibinfo{pages}{053003} (\bibinfo{year}{1999}), \eprint{hep-ph/9903554}.

\bibitem[{\citenamefont{Miranda et~al.}(2006)\citenamefont{Miranda, Tortola,
  and Valle}}]{Miranda:2004nb}
\bibinfo{author}{\bibfnamefont{O.}~\bibnamefont{Miranda}},
  \bibinfo{author}{\bibfnamefont{M.}~\bibnamefont{Tortola}}, \bibnamefont{and}
  \bibinfo{author}{\bibfnamefont{J.~W.~F.}~\bibnamefont{Valle}},
  \bibinfo{journal}{JHEP} \textbf{\bibinfo{volume}{0610}},
  \bibinfo{pages}{008} (\bibinfo{year}{2006}), \eprint{hep-ph/0406280}.

\bibitem[{\citenamefont{Bolanos et~al.}(2009)\citenamefont{Bolanos, Miranda, Palazzo,
Tortola, and Valle}}]{Bolanos:2008km}
\bibinfo{author}{\bibfnamefont{A.}~\bibnamefont{Bolanos}},
  \bibinfo{author}{\bibfnamefont{O.}~\bibnamefont{Miranda}},
  \bibinfo{author}{\bibfnamefont{A.}~\bibnamefont{Palazzo}},
  \bibinfo{author}{\bibfnamefont{M.}~\bibnamefont{Tortola}}, \bibnamefont{and}
  \bibinfo{author}{\bibfnamefont{J.~W.~F.}~\bibnamefont{Valle}},
  \bibinfo{journal}{Phys.~Rev.} \textbf{\bibinfo{volume}{D79}},
  \bibinfo{pages}{113012} (\bibinfo{year}{2009}), \eprint{0812.4417}.

\bibitem[{\citenamefont{Escrihuela et~al.}(2009)\citenamefont{Escrihuela, Miranda,
Tortola, and Valle}}]{Escrihuela:2009up}
\bibinfo{author}{\bibfnamefont{F.}~\bibnamefont{Escrihuela}},
  \bibinfo{author}{\bibfnamefont{O.}~\bibnamefont{Miranda}},
  \bibinfo{author}{\bibfnamefont{M.}~\bibnamefont{Tortola}}, \bibnamefont{and}
  \bibinfo{author}{\bibfnamefont{J.~W.~F.}~\bibnamefont{Valle}},
  \bibinfo{journal}{Phys.~Rev.} \textbf{\bibinfo{volume}{D80}},
  \bibinfo{pages}{105009} (\bibinfo{year}{2009}), \eprint{0907.2630}.

\bibitem[{\citenamefont{Fornengo et~al.}(2002)\citenamefont{Fornengo, Maltoni,
Tomas, and Valle}}]{Fornengo:2001pm}
\bibinfo{author}{\bibfnamefont{N.}~\bibnamefont{Fornengo}},
  \bibinfo{author}{\bibfnamefont{M.}~\bibnamefont{Maltoni}},
  \bibinfo{author}{\bibfnamefont{R.}~\bibnamefont{Tomas}}, \bibnamefont{and}
  \bibinfo{author}{\bibfnamefont{J.~W.~F.}~\bibnamefont{Valle}},
  \bibinfo{journal}{Phys.~Rev.} \textbf{\bibinfo{volume}{D65}},
  \bibinfo{pages}{013010} (\bibinfo{year}{2002}), \eprint{hep-ph/0108043}.

\bibitem[{\citenamefont{Huber and Valle}(2001)}]{Huber:2001zw}
\bibinfo{author}{\bibfnamefont{P.}~\bibnamefont{Huber}} \bibnamefont{and}
  \bibinfo{author}{\bibfnamefont{J.~W.~F.} \bibnamefont{Valle}},
  \bibinfo{journal}{Phys.~Lett.} \textbf{\bibinfo{volume}{B523}},
  \bibinfo{pages}{151} (\bibinfo{year}{2001}), \eprint{hep-ph/0108193}.

\bibitem[{\citenamefont{Mitsuka et~al.}(2011)}]{Mitsuka:2011ty}
\bibinfo{author}{\bibfnamefont{G.}~\bibnamefont{Mitsuka}} \bibnamefont{et~al.}
  (\bibinfo{collaboration}{Super-Kamiokande Collaboration}),
  \bibinfo{journal}{Phys.~Rev.} \textbf{\bibinfo{volume}{D84}},
  \bibinfo{pages}{113008} (\bibinfo{year}{2011}), \eprint{1109.1889}.

\bibitem[{\citenamefont{Escrihuela et~al.}(2011)\citenamefont{Escrihuela, Tortola,
Valle, and Miranda}}]{Escrihuela:2011cf}
\bibinfo{author}{\bibfnamefont{F.}~\bibnamefont{Escrihuela}},
  \bibinfo{author}{\bibfnamefont{M.}~\bibnamefont{Tortola}},
  \bibinfo{author}{\bibfnamefont{J.~W.~F.}~\bibnamefont{Valle}}, \bibnamefont{and}
  \bibinfo{author}{\bibfnamefont{O.}~\bibnamefont{Miranda}},
  \bibinfo{journal}{Phys.~Rev.} \textbf{\bibinfo{volume}{D83}},
  \bibinfo{pages}{093002} (\bibinfo{year}{2011}), \eprint{1103.1366}.

\bibitem[{\citenamefont{Ohlsson et~al.}(2013)\citenamefont{Ohlsson, Zhang,
and Zhou}}]{Ohlsson:2013epa}
\bibinfo{author}{\bibfnamefont{T.}~\bibnamefont{Ohlsson}},
  \bibinfo{author}{\bibfnamefont{H.}~\bibnamefont{Zhang}}, \bibnamefont{and}
  \bibinfo{author}{\bibfnamefont{S.}~\bibnamefont{Zhou}},
  \bibinfo{journal}{Phys.~Rev.} \textbf{\bibinfo{volume}{D88}},
  \bibinfo{pages}{013001} (\bibinfo{year}{2013}), \eprint{1303.6130}.

\bibitem[{\citenamefont{Esmaili and Smirnov}(2013)}]{Esmaili:2013fva}
\bibinfo{author}{\bibfnamefont{A.}~\bibnamefont{Esmaili}} \bibnamefont{and}
  \bibinfo{author}{\bibfnamefont{A.~Y.} \bibnamefont{Smirnov}},
  \bibinfo{journal}{JHEP} \textbf{\bibinfo{volume}{1306}},
  \bibinfo{pages}{026} (\bibinfo{year}{2013}), \eprint{1304.1042}.

\bibitem[{\citenamefont{Blennow et~al.}(2008)\citenamefont{Blennow, Ohlsson,
and Skrotzki}}]{Blennow:2007pu}
\bibinfo{author}{\bibfnamefont{M.}~\bibnamefont{Blennow}},
  \bibinfo{author}{\bibfnamefont{T.}~\bibnamefont{Ohlsson}}, \bibnamefont{and}
  \bibinfo{author}{\bibfnamefont{J.}~\bibnamefont{Skrotzki}},
  \bibinfo{journal}{Phys.~Lett.} \textbf{\bibinfo{volume}{B660}},
  \bibinfo{pages}{522} (\bibinfo{year}{2008}), \eprint{hep-ph/0702059}.

\bibitem[{\citenamefont{Ribeiro et~al.}(2007)\citenamefont{Ribeiro, Minakata,
Nunokawa, Uchinami, and Zukanovich-Funchal}}]{Ribeiro:2007ud}
\bibinfo{author}{\bibfnamefont{N.}~\bibnamefont{Ribeiro}},
  \bibinfo{author}{\bibfnamefont{H.}~\bibnamefont{Minakata}},
  \bibinfo{author}{\bibfnamefont{H.}~\bibnamefont{Nunokawa}},
  \bibinfo{author}{\bibfnamefont{S.}~\bibnamefont{Uchinami}}, \bibnamefont{and}
  \bibinfo{author}{\bibfnamefont{R.}~\bibnamefont{Zukanovich-Funchal}},
  \bibinfo{journal}{JHEP} \textbf{\bibinfo{volume}{0712}},
  \bibinfo{pages}{002} (\bibinfo{year}{2007}), \eprint{0709.1980}.

\bibitem[{\citenamefont{Blennow and Ohlsson}(2008)}]{Blennow:2008eb}
\bibinfo{author}{\bibfnamefont{M.}~\bibnamefont{Blennow}} \bibnamefont{and}
  \bibinfo{author}{\bibfnamefont{T.} \bibnamefont{Ohlsson}},
  \bibinfo{journal}{Phys.~Rev.} \textbf{\bibinfo{volume}{D78}},
  \bibinfo{pages}{093002} (\bibinfo{year}{2008}), \eprint{0805.2301}.

\bibitem[{\citenamefont{Adamson et~al.}(2013)}]{Adamson:2013ovz}
\bibinfo{author}{\bibfnamefont{P.}~\bibnamefont{Adamson}} \bibnamefont{et~al.}
  (\bibinfo{collaboration}{MINOS Collaboration}),
  \bibinfo{journal}{Phys.~Rev.} \textbf{\bibinfo{volume}{D88}},
  \bibinfo{pages}{072011} (\bibinfo{year}{2013}), \eprint{1303.5314}.

\bibitem[{\citenamefont{Kopp et~al.}(2007)\citenamefont{Kopp, Lindner,
and Ota}}]{Kopp:2007mi}
\bibinfo{author}{\bibfnamefont{J.}~\bibnamefont{Kopp}},
  \bibinfo{author}{\bibfnamefont{M.}~\bibnamefont{Lindner}}, \bibnamefont{and}
  \bibinfo{author}{\bibfnamefont{T.}~\bibnamefont{Ota}},
  \bibinfo{journal}{Phys.~Rev.} \textbf{\bibinfo{volume}{D76}},
  \bibinfo{pages}{013001} (\bibinfo{year}{2007}), \eprint{hep-ph/0702269}.

\end{thebibliography}
\end{document}